\documentclass[pra,showpacs,preprint]{revtex4} \usepackage{latexsym}
\usepackage{amsfonts} \usepackage{amsmath,revsymb,graphicx} 

\newcommand{\kt}[1]{\ensuremath{|#1\rangle}}

\newcommand{\ckt}[1]{\ensuremath{|#1\}}}
\newcommand{\cbr}[1]{\ensuremath{\{ #1|}}
\newcommand{\cbk}[2]{\ensuremath{\{ #1|#2\}}}
\newcommand{\pkt}[1]{\ensuremath{\left|#1\right)}}

\begin{document}

\title[Symmetries of Three Harmonically-Trapped Particles in One Dimension]{Symmetries of Three Harmonically-Trapped Particles in One Dimension}

\author{N.L.~Harshman\footnote{Electronic address: harshman@american.edu}}

\affiliation{Department of Physics\\
4400 Massachusetts Ave. NW\\ American University\\ Washington, DC 20016-8058}

\begin{abstract}
We present a method for solving trapped few-body problems and apply it to three equal-mass particles in a one-dimensional harmonic trap, interacting via a contact potential.  By expressing the relative Hamiltonian in Jacobi cylindrical coordinates, i.e.\ the two-dimensional version of three-body hyperspherical coordinates, we discover an underlying ${\rm C}_{6v}$ symmetry.  This symmetry simplifies the calculation of energy eigenstates of the full Hamiltonian in a truncated Hilbert space constructed from the trap Hamiltonian eigenstates.  Particle superselection rules are implemented by choosing the relevant representations of ${\rm C}_{6v}$.  We find that the one-dimensional system shows nearly the full richness of the three-dimensional system, and can be used to understand separability and reducibility in this system and in standard few-body approximation techniques.
\end{abstract}

\maketitle

\section{Introduction}

The advent of precision control and measurement of interacting ultracold atomic systems has provided a productive testing ground for few-body physics.  Predictions from condensed matter and nuclear physics made in the 1960's and 1970's, like fermionization of strongly-interacting bosons and universality in Efimovian states, have been verified using ultracold atoms as a kind of quantum simulator.  The signatures of few body physics can also be seen in the statistical and dynamical properties of ultracold quantum gases.  Further, deterministic loading of a few cold atoms into deep optical lattices has made feasible the prospect of spectroscopic measurements of highly-correlated atomic states.

This last application in particular has generated a lot of interest in analytical results and theoretical methods for a few particles that are trapped in an approximately harmonic well and interacting via short-range potentials~\cite{busch_two_1998,blume_three_2002,jonsell_interaction_2002,jonsell_universal_2002,tan_short_2004, stoll_production_2005,wang_s-wave_2005, werner_unitary_2006-1,werner_unitary_2006,luu_three-fermion_2007,kestner_level_2007,stetcu_effective_2007,gogolin_analytic_2008,stetcu_effective_2010,tolle_spectrum_2010,daily_energy_2010,rotureau_three_2010,tolle_universal_2011,armstrong_analytic_2011,portegies_efimov_2011,grishkevich_theoretical_2011,zinner_universal_2012,johnson_effective_2012}. See \cite{blume_few-body_2012} for a recent review and a more complete bibliography, including experimental progress.    One motivation for this line of research  is, like untrapped few-body systems, trapped systems offer the possibility for observing universal properties~\cite{greene_universal_2010,zinner_universal_2012}, i.e.\ few-body observables whose value is independent of the short-range interaction details or particle structure~\cite{braaten_universality_2006}.

For harmonic traps, exact results exist for two bodies in the limit of a zero-range, contact interaction~\cite{busch_two_1998,jonsell_interaction_2002}, sometimes called the scaling limit.  For three harmonically-trapped particles, there are analytic solutions in the joint case of the scaling limit combined with the unitary (or resonance) limit~\cite{jonsell_universal_2002,werner_unitary_2006-1,gogolin_analytic_2008}, i.e\ strong interactions, usually described by a large two-body scattering length. However, except for the unitary limit, solving for the energies and eigenstates of few-body quantum systems requires approximations.  This article describes a method to analyze the case of three equal-mass particles in one-dimension interacting via a zero-range contact interaction.  We will calculate the variation of the energy spectrum as a function of interaction strength as it varies between the repulsive unitary limit to the attractive unitary limit.  We will combine techniques that have have been previously employed with a geometrical approach to implementing the discrete symmetries of the Hamiltonian and any superselection rules due to indistinguishable particles. In particular, we find that representations of the symmetry groups $\rm{C}_{6v}$ and  $\rm{C}_{2v}$ allow us to present a unified classification scheme for the atom, dimer and trimer states with different three-particle content, including two and three identical bosons or fermions, as well as no identical particles.  This is a generalizable method that maximizes separability and reducibility in trapped few-body systems and therefore enhances computational efficiency.  Our method is most similar in spirit to the techniques of \cite{kilpatrick_set_1987}, who use discrete symmetries to classify three-particle states in two dimensions, and \cite{grishkevich_theoretical_2011}, who use discrete symmetries to reduce the problem of two-particles in an anisotropic well.

\subsection{Separability}

Approximations are particularly useful when they exploit symmetries to map the few-body system in a controlled way onto a related mathematical structure that has more separability and reducibility.  
The standard example of separability is found in the transformation from the particle coordinates to the center-of-mass/relative coordinates, e.g.\ Jacobi coordinates, allowing the energy eigenstates to be expressed as an unentangled product of center-of-mass and relative wave functions.  
This separation is exact for any configuration of free particles due to Galilean invariance, and it is also exact for a quadratic trap (even with unequal masses) as long as the non-interacting trap oscillation frequencies for the particles are all the same~\cite{werner_unitary_2006}.  Here we will restrict ourselves to the case of equal masses for simplicity.  This will ensure center-of-mass vs.\ relative separability even when we add interparticle interactions to construct the full Hamiltonian.  The case of equal masses also provides another kind of separability.  As we discuss below, the trap Hamiltonian has ${\rm SU}(3)$ symmetry, and this symmetry allows the trap Hamiltonian to be superintegrable, and therefore separable in multiple coordinate systems.

Beyond these exact separability results, a standard class of approximations to increase separability are called adiabatic approximations.  For example, the Born-Oppenheimer approximation effectively separates the electronic and nuclear wave functions by assuming the electronic wave function varies at a much faster scale than the nuclear wave function.  In the context of the three-body problem, a similar approximation is made when using hyperspherical coordinates to represent the relative degrees of freedom when solving time-independent Schr\"odinger equation~\cite{blume_three_2002,werner_unitary_2006-1,daily_energy_2010}  or Faddeev equations~\cite{fedorov_efimov_1993,jonsell_universal_2002} in coordinate space.  A good low-energy approximation is to effectively separate the wave function in terms of the hyperradius and hyperangle by neglecting the variation of the wave function in hyperradius when solving the hyperangular equation.  This adiabatic approximation becomes exact in the unitary limit~\cite{jonsell_universal_2002}.  This occurs because the zero-range contact interaction preserves a dynamical scale invariance that introduces a hidden $\rm{SO}(2,1)$ symmetry to the problem~\cite{werner_unitary_2006}.

An advantage of working in one-dimension is that the six-dimensional `three-body hyperspherical relative coordinates' are just two-dimensional cylindrical coordinates.  In one-dimension, the separation between hyperangle and hyperradius that emerges in the unitary limit will be easily evident; the three-fermion states are the exact solution in this limit and we will see that they are separable in this sense.

\subsection{Reducibility}

The truncated Hilbert space approximation maps the infinite-dimensional Hilbert space of the trapped three-body system onto a truncated model space constructed from exact solutions of the trap Hamiltonian without particle interactions.  In a sense, this employs the most basic sense of reducibility: all finite-dimensional Hilbert spaces can be reduced to a sum of subspaces corresponding to degenerate eigenvalues of the full Hamiltonian.  Procedures can be used to generate an effective Hamiltonian that is  corrected for the energy cut-off due to truncation and which handle the difficulties in implementing a possibly non-Hermitian, non-renormalizaible, highly singular zero-range interparticle interaction in a consistent way.  For example, such a method has employed by the no-core shell model with a Hamiltonian renormalized using effective field theory~\cite{navratil_few-nucleon_2000, stetcu_no-core_2007,stetcu_effective_2007,rotureau_three_2010}.  Similar truncated Hilbert space models but with different approaches to regularization are employed in \cite{luu_three-fermion_2007} and \cite{tolle_spectrum_2010,tolle_universal_2011}.  In one-dimension, we will not have these difficulties with renormalization and regularization, but we will be able to clearly see the effect of the energy cut-off on the deviation of approximate results from the exact values in the universal, unitary limit.

To further enhance reducibility, we will exploit the discrete point symmetry ${\rm C}_{6v}$ of the relative configuration space.  This symmetry group is the geometrical realization of the permutation group of three particles, combined with the overall parity inversion symmetry.  Utilizing properties of the irreducible representations of this group make two calculational steps much more efficient, as is well-known in applications from molecular and chemical physics.  First, diagonalizing the full Hamiltonian in the truncated Hilbert space can be made more efficient because there are only non-zero matrix elements of the interaction between states in the same representation of ${\rm C}_{6v}$.  Second, implementing superselection rules to handle indistinguishable particles is automatic: only a few representations are relevant depending on which particles (if any) are indistinguishable.  We will apply our method to the cases of three indistinguishable bosons (BBB), two bosons and another particle (BBX), three indistinguishable fermions (FFF), two indistinguishable fermions and one other particle (FFX) and three distinguishable particles (XYZ).  As discussed below, we assume there is no entanglement between the internal states of the particles and the motional state, or among the internal states of the particle.  It is specifically the application of the ${\rm C}_{6v}$ symmetry, when used in combination with hyperspherical coordinates and a truncated Hilbert space approach, that we argue provides a meaningful efficiency improvement over existing methods.  In particular, several authors have discussed the computational difficulties of implementing particle symmetries in a variety of truncated Hilbert space models~\cite{navratil_few-nucleon_2000, rotureau_three_2010,tolle_spectrum_2010,daily_energy_2010}, and we believe this is an elegant, extensible method for solving these difficulties efficiently.

\subsection{Motivation}

Besides our motivation to demonstrate our method and its connection to previous methods, there is genuine theoretical and experimental interest in quantum gases and other ultracold atomic systems trapped in one-dimension.  See, for example, the recent review \cite{cazalilla_one_2011}. Therefore, we do not consider this application to be merely a ``toy model''.  An effective one-dimensional description emerges as a limit of the three-dimensional case as the harmonic trap becomes long in one direction and tight in the transverse plane, see for example~\cite{lieb_one-dimensional_2003}.  A surprising number of the structures of interest that are found in the three-dimensional case, universal and otherwise,  will also be manifest in the one-dimensional problem~\cite{gangardt_universal_2004}.  For example, when the interaction is attractive and sufficiently strong for three indistinguishable bosons, the system manifests all the classes of bound states found for attractive contact interactions in a three-dimensional harmonic potential: trimer states, dimer plus single atom states, and states of three bound atoms. In the spectra below, the emergence of universality in the unitary limit also will be evident.   

In the next section, we describe the model Hamiltonian for three equal-mass particles in one-dimension, establish notation, and begin a discussion of the symmetries of the Hamiltonian that will continue into Section III.  In both sections, the usefulness of the Jacobi hyperspherical (i.e.\ cylindrical) relative coordinates for visualizing and analyzing configuration space will be emphasized. In Section IV, we provide results for the energy eigenstates and show how the richness of three-body physics is evidenced even in this simple model.  Beyond the advantages of analytic power and explanatory appeal, studying this system in one-dimension also suggests extensions of this method to more complicated cases, like more particles, non-equal mass particles, genuine multi-body interactions, higher dimensions, and anisotropic traps.  The final section contains a few remarks about these possible applications.

\section{The Model}

The total Hamiltonian $H$ can be written as the sum of the harmonic trap Hamiltonian $H_0$, which in position space is
 \begin{equation}
H_0 = \frac{1}{2m}\left(\tilde{p}_1^2+\tilde{p}_2^2+\tilde{p}_3^2\right) + \frac{m\omega^2}{2}\left(\tilde{q}_1^2+\tilde{q}_2^2+\tilde{q}_3^2\right)
\label{eq:H0unitc}
\end{equation}
and $V$, the sum of the three two-particle interaction terms
\begin{equation}
V= g \left(\delta(\tilde{q}_1 - \tilde{q}_2) + \delta(\tilde{q}_2 - \tilde{q}_3)+ \delta(\tilde{q}_3 - \tilde{q}_1)\right),
\label{eq:Vunit}
\end{equation}
where $\tilde{p}_j = (\hbar/i) \partial/\partial \tilde{q}_j$.  The potential $V$ represents the idealization of a zero-range `contact interaction'.  Because it is comprised of the sum of three one-dimensional delta functions and it is self-adjoint, there will be no need for regularization, renormalization or other techniques to tame singularities and divergences the full Hamiltonian in the truncated shell model (although such techniques can improve the rate of convergence).

As is usual, we change to scaled, unitless coordinates using the length scale $\sigma=\sqrt{\hbar/m\omega}$:
\begin{equation}
\tilde{q}_i= q_i \sigma\ \mbox{and}\ \tilde{p}_i=  p_i \hbar/\sigma.
\label{eq:scale}
\end{equation}
Then defining ${\bf p}=(p_1,p_2,p_3)^\top$ and ${\bf q}=(q_1,q_2,q_3)^\top$, the Hamiltonian takes the simpler form
\begin{eqnarray}
H_0 &=&  \frac{\hbar \omega}{2}\left({\bf p}^2 + {\bf q}^2\right)\label{eq:H0part}\\
V&=&\frac{g}{\sigma}\left(\delta(q_1 - q_2) + \delta(q_2 - q_3)+ \delta(q_3 - q_1)\right).\label{eq:Vpart}
\end{eqnarray}

The form (\ref{eq:H0part}) for $H_0$  suggests the trivial observation that the Hamiltonian of three equal-mass harmonic oscillators in one dimension is isomorphic to one isotropic harmonic oscillator in three dimensions (forgetting for the moment about any superselection rules).  This system and its $\rm{SU}(3)$ symmetry are well-known in the literature.  Here we note that the trap Hamiltonian is separable in cartesian, cylindrical and spherical coordinates and therefore we can define three classes of energy eigenbasis vectors:
\begin{eqnarray}
\mbox{cartesian:}\ H_0 \kt{n_1, n_2 ,n_3} &=& \hbar\omega (n_1 + n_2 + n_3 + 3/2) \kt{n_1, n_2, n_3}\nonumber\\
\mbox{cylindrical:}\ H_0 \ckt{n_\rho, m ,n_3} &=& \hbar\omega (2n_\rho + |m| + n_3 + 3/2) \ckt{n_\rho, m, n_3}\nonumber\\
\mbox{spherical:}\ H_0 \pkt{n_r, \ell, \ell_3} &=& \hbar\omega (2n_r + \ell + 3/2) \pkt{n_r, \ell ,\ell_3}.
\label{eq:3bases}
\end{eqnarray}
In these equations, all the $n$'s and $\ell$ can be any non-negative integers and $m$ and $\ell_3$ can be any integer.  We define the total excitation number $N$ as
\begin{equation}
N=n_1+n_2+n_3 = 2n_\rho + |m| + n_3 = 2n_r + \ell
\label{eq:totalN}
\end{equation}
In the absence of any superselection rules, the eigenspace associated to energy $E_N = \hbar\omega(N+3/2)$ has a dimension of $d_N = 1/2 (N+1)(N+2)$.  The three classes of bases in (\ref{eq:3bases}) are just different ways of diagonalizing the degeneracies of $\mathcal{H}_N$, the $E_N$ subspace of $\mathcal{H}$.  From this we can also calculate the dimension of the truncated Hilbert space $\sum_{i=0}^N \mathcal{H}_i$ with maximum excitation number $N$ as  $1/6(N+1)(N+2)(N+3)$.

For each of the three types of separable bases in (\ref{eq:3bases}), the ${\rm SU(3)}$ symmetry also provides an equivalence class of energy eigenbases.  Take $U(u)$ as a unitary representation of $u\in{\rm SU}(3)$, then any bases of the form
\begin{eqnarray}
\mbox{cartesian:}\ \kt{n_1 ,n_2 ,n_3}_u&\equiv& U(u)\kt{n_1 ,n_2, n_3}\nonumber\\
\mbox{cylindrical:}\ \ckt{n_\rho ,m, n_3}_u &\equiv& U(u)\ckt{n_\rho ,m ,n_3}\nonumber\\
\mbox{spherical:}\ \pkt{\nu_r, \ell ,\ell_3}_u &\equiv& U(u)\pkt{n_r, \ell ,\ell_3}
\label{eq:3ubases}
\end{eqnarray}
are also complete energy eigenbases for the total Hilbert space $\mathcal{H}$.  Further, $U(u)$ only connects basis vectors with the same total excitation $N$ (note that $U(u)$ may be reducible within the degenerate energy eigenspace $\mathcal{H}_N$).  As an example, in the cartesian basis $\kt{n_1 ,n_2 ,n_3}_u$, the quantum numbers $n_i$ no longer represent particle excitations, but excitations of some state corresponding to a particular structure of particle correlations characterized by the particular element $u$.

\begin{figure}
\centering
\includegraphics[width=0.5\linewidth]{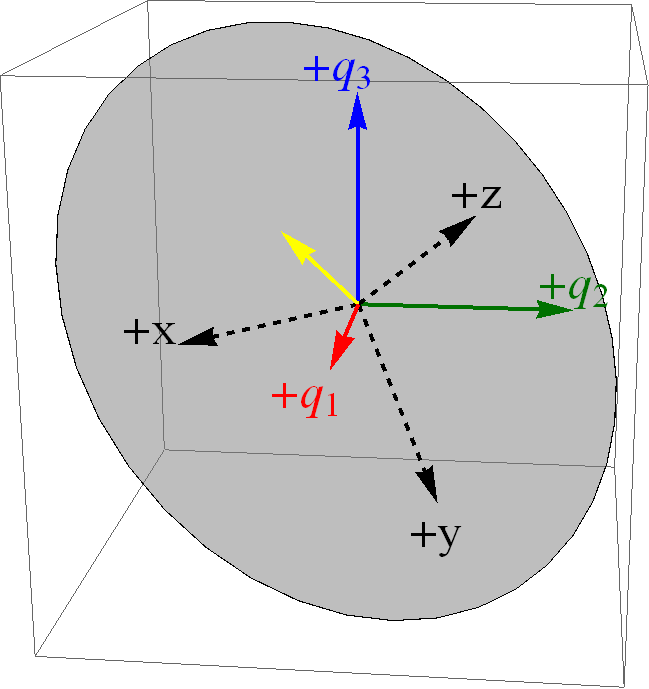}
\caption{(Color online) This figure demonstrates that the transformation to Jacobi coordinates can be represented as a rotation in configuration space.  The $\hat{q}_i$ positive particle axes are solid arrows; The Jacobi cartesian positive axes are dashed arrows.  The $\hat{z}$ direction points into the middle of the solid angle bounded by the all-positive octant and the $\hat{x}$ and $\hat{y}$ are orthogonal directions in the relative plane, depicted as a gray disk for $z=0$, i.e.\ the center-of-mass at the bottom of the potential well.  This Jacobi rotation takes place around the yellow (light gray, unlabeled) vector and is about 1.22 radians.}
\label{fig1} 
\end{figure}

Consider the specific element $J\in{\rm SO}(3)\subset{\rm SU}(3)$ corresponding to the rotation in configuration space from particle coordinates ${\bf q}$ to one set of Jacobi coordinates $J{\bf q}={\bf r}=(x,y,z)^\top$:
\begin{equation}
J =\left(\begin{array}{ccc}
\frac{1}{\sqrt{2}}& -\frac{1}{\sqrt{2}}&0 \\ \frac{1}{\sqrt{6}}& \frac{1}{\sqrt{6}}& -\frac{2}{\sqrt{3}}\\  \frac{1}{\sqrt{3}} & \frac{1}{\sqrt{3}} & \frac{1}{\sqrt{3}} \end{array}\right).
\label{eq:Jtrans}
\end{equation}
Coordinate $z= 1/\sqrt{3}(q_1 +q_2 + q_3)$ is the (normalized) center-of-mass of three particles and coordinates $x=1/\sqrt{2}(q_1 - q_2)$ and $y=1/\sqrt{6}(q_1+q_2) -\sqrt{2/3}q_3$ are one possible arrangement of the standard (normalized) Jacobi coordinates~\footnote{Some authors use a non-orthonormal transformation for Jacobi cartesian coordinates, and then use extra numerical factors when, for example, converting to cylindrical (hyperspherical) coordinates.  We prefer the orthonormal definition so that $J^{-1}=J^\top$ and therefore the momentum transforms canonically under the same rotation and there is no need to adjust mass values.  Additionally, some use the definition $\tan\alpha=x/y$ (compare to (\ref{eq:Jcyl})).}.  In these coordinates, the Hamiltonian becomes
\begin{eqnarray}
H_0 &=&  \frac{\hbar \omega}{2}\left({\bf k}^2 + {\bf r}^2\right)\label{eq:H0config}\\
V&=&\frac{g}{\sqrt{2}\sigma}\left(\delta\left(x\right) + \delta\left(-{\scriptstyle \frac{1}{2}}x + {\scriptstyle \frac{\sqrt{3}}{2}}y\right)+ \delta\left(-{\scriptstyle \frac{1}{2}}x - {\scriptstyle \frac{\sqrt{3}}{2}}y\right)\right),\label{eq:Vconfig}
\end{eqnarray}
where ${\bf k}=J{\bf p}$.  See Figure \ref{fig1} for a graphical representation of the Jacobi rotation $J$. Note that since $V$ does not depend on $z$, the full Hamiltonian still separates in terms of center-of-mass and relative coordinates.  This means that instead of needing $1/6(N+1)(N+2)(N+3)$ states to calculate the ground state energy in a truncated Hilbert space with maximum energy $E_N$, we only need $1/2(N+1)(N+2)$ because there are only non-zero matrix elements of $V$ between vectors with the same quantum number for center-of-mass excitations.

\begin{figure}
\centering
\includegraphics[width=0.5\linewidth]{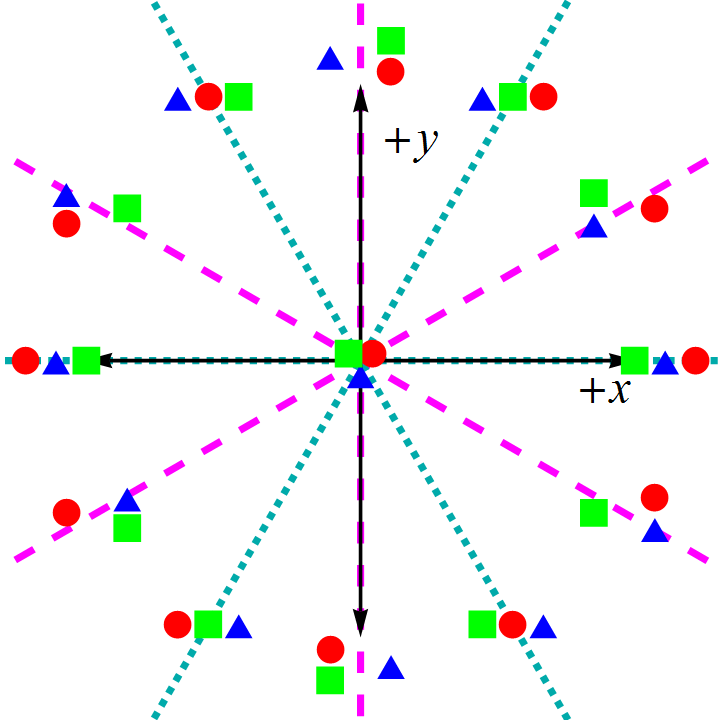}
\caption{(Color online) This figure depicts the relative plane in Jacobi coordinates.  The magenta (dashed) lines have several interpretations.  First, they are the projections of the particle axes on the relative plane (see Figure \ref{fig1}).  For example, the line with $\varphi=\pi/6$ corresponds to increasing separation of particle 1 (depicted as a red circle) from  particles 2 (green square) and 3 (blue triangle), with particle 1 on positive side and the other two co-located on the negative side.  Reflections across the magenta (dashed) lines also represent two-particle exchanges, e.g. the reflection through the line $\varphi=\pi/6$ corresponds to the $\hat{\sigma}_{23}$ 2-cycle.  Rotations of $\pm 2\pi/3$ are 3-cycles, i.e.\ three-particle exchanges.  A rotation of $\pi$ corresponds to a reversal of parity $\hat{\pi}$.  Reflections across the cyan (dotted) lines are a combination of a two-particle exchange and a parity inversion, e.g. $\hat{\pi}\hat{\sigma}_{23}$.}
\label{fig2} 
\end{figure}

We will denote the cylindrical $H_0$ energy eigenbasis in the Jacobi coordinate system as $\ckt{\nu, \mu, \eta} \equiv \ckt{n_\rho ,m, n_3}_J$.  In the next section, we will show that these states are a convenient basis for irreducible representations of the ${\rm C}_{6v}$ symmetry of the full Hamiltonian $H$, but here we note a few additional properties.  The center-of-mass motion separates from the relative motion so $\ckt{\nu ,\mu, \eta}=\ckt{\nu ,\mu}\otimes\ckt{ \eta}$.  The wave functions realizing the center-of-mass state $\ckt{\eta}$ in the $z$-coordinate  are the normal one-dimensional harmonic oscillator states:
\begin{equation}
\cbk{z}{\eta} = \psi_\eta(z) = \pi^{-1/4} \left(2^\eta \eta!\right)^{-1/2} H_\eta (z) \exp\left(-\frac{1}{2}z^2\right).
\label{eq:1DHO}
\end{equation}
On the plane perpendicular to $z$ that describes the relative motion of the three particles, we can define the ``hyperspherical coordinates'' $\rho$ and $\varphi$ from $x$ and $y$ using the standard conversion to cylindrical coordinates
\begin{eqnarray}
\rho&=&\sqrt{x^2 + y^2} = \sqrt{q_1^2 + q_2^2 + q_3^2 -q_1 q_2 - q_2 q_3 -q_3 q_1}\ \mbox{and}\nonumber\\
\tan\varphi&=&y/x = \frac{ q_1 + q_2 - 2 q_3}{\sqrt{3} q_1 - \sqrt{3} q_2}.
\label{eq:Jcyl}
\end{eqnarray}
Then the realization of the relative state $\ckt{\nu, \mu}$ as wave functions in the Jacobi cylindrical coordinate $\rho$ and $\varphi$ are
\begin{equation}
\cbk{\rho,\varphi}{\nu,\mu} = \psi_{\nu,\mu}(\rho,\varphi) = \sqrt{\frac{\nu!}{\pi (\nu+|\mu|)!}}\rho^{|\mu|} L^{|\mu|}_{\nu}(\rho^2)e^{-\rho^2/2}e^{i\mu\varphi}\label{eq:Jcylwave}.
\end{equation}
Note that only $\mu=0$ states are not zero at $\rho=0$ and therefore have some probability density for configurations in which all three particles are located in the same place.  
See Figure \ref{fig2} for further explanation of the correspondence between particle configurations and Jacobi cylindrical coordinates.

\section{Symmetries and the Superselection Rule}

Expressed in Jacobi cylindrical coordinates, the interaction $V$ takes takes the form~\footnote{The apparent singularity at $\rho=0$ in this equation does not cause problems for the integrals we will be considering because of the measure of integration will cancel this singularity.}
\begin{equation}
V = \frac{g}{\sqrt{2}\sigma \rho}\sum_{i=1}^6\delta\left(\varphi - {\scriptstyle\frac{(2i-1)}{6}}\pi\right).
\label{eq:Vpolar}
\end{equation}
To understand this functional form, note that the zeros of the delta-functions define three planes in full configuration space that intersect the relative $xy$-plane in three lines (the dashed magenta lines in Figure \ref{fig2}).  The locus of points on these planes correspond to configurations with two particles in the same position.  The rays along $\varphi=\pi/6$, $5\pi/6$, and $3\pi/2$ correspond to the configurations where one particle has a positive $q$-coordinate and the other two  particles have the same negative $q$-coordinate.  The rays with angles $\varphi=\pi/2$, $7\pi/6$, and $11\pi/6$ are configurations with the reverse orientation.

From this form we can see that the potential $V$ manifests the two-dimensional point symmetry that is usually denoted ${\rm C}_{6v}$, i.e.\ the symmetries of the regular hexagon.  The finite group ${\rm C}_{6v}$ has twelve elements: the identity, two rotations by $\pi/3$, two rotations by $2\pi/3$, one rotation by $\pi$ and six reflections of two different types.  The group ${\rm C}_{6v}$ is isomorphic to ${\rm S}_3 \times {\rm Z}_2$, where ${\rm S}_3$ is the permutation group of three objects and would be geometrically realized the Jacobi relative plane by ${\rm C}_{3v}$.  The ${\rm Z}_2$ symmetry corresponds to the parity invariance of the relative configuration space, which is respected by both $H_0$ and $V$\footnote{One could also include another factor of ${\rm Z}_2$ to realize the parity invariance of the Hamiltonian in the center-of-mass coordinate $z$ and then the geometric realization would be $D_{6h}$, the point symmetry of a regular hexagonal prism.  This would cause an additional doubling of representations, but since the interaction does not affect the center-of-mass it does not provide any additional efficiency in calculations.}.  We list the group elements and how they act on the relative plane and transform the Jacobi coordinates in  Table \ref{tab:elements}.

All ${\rm C}_{6v}$ group elements leave the hyperradial coordinate $\rho$ unchanged, but they transform $\varphi\rightarrow \varphi'$.  This connection can be inferred from the the geometry of Figure  \ref{fig2}, or more systematically by using the matrix $J$ to transform the relevant operator.  For example, in the cartesian particle basis, the three-particle exchange $1\rightarrow 2$, $2\rightarrow 3$ and $3\rightarrow 1$, denoted $\hat{\sigma}_{231}$  is represented in particle coordinate ${\bf q}$-space as
\begin{equation}
\hat{\sigma}_{231} \rightarrow S_{231} = \left( \begin{array}{ccc} 0 & 1 &0 \\ 0 & 0 & 1 \\ 1 & 0& 0\end{array}\right).
\label{eq:12mat}
\end{equation}
The representation in terms of Jacobi coordinates ${\bf r}$-space can be found by transformation the matrix with $J$:
\begin{equation}
JS_{231}J^\top = \left( \begin{array}{ccc}\scriptstyle  -\frac{1}{2} & \scriptstyle -\frac{\sqrt{3}}{2} &0 \\ \scriptstyle \frac{\sqrt{3}}{2} & \scriptstyle -\frac{1}{2} & 0 \\ 0 & 0& 1 \end{array}\right),
\label{eq:12matJ}
\end{equation}
which can be recognized as a rotation around the center-of-mass $\hat{z}$-direction by $-2\pi/3$.

\begin{table}
\centering
\begin{tabular}{|c|c|c|}
\hline
$g\in{\rm C}_{6v}$ & $g\in{\rm S}_3 \times {\rm Z}_2$ & $\varphi\rightarrow \varphi'$ \\
\hline
$E$ & $\hat{e}$  & $\varphi$ \\
$\sigma_{v}$ & $\hat{\sigma}_{12}$ & $-\varphi + \pi$\\
$\sigma_{v'} $ & $\hat{\sigma}_{23} $ & $-\varphi + \frac{\pi}{3}$ \\
$\sigma_{v''} $ & $\hat{\sigma}_{31} $ & $-\varphi - \frac{\pi}{3}$ \\
$C_3^{-1}$ & $\hat{\sigma}_{231}$ & $\varphi - \frac{2\pi}{3}$ \\
$C_3$ & $\hat{\sigma}_{312}$ & $\varphi + \frac{2\pi}{3}$ \\
$C_2$ & $\hat{\pi}$  & $\varphi+\pi$ \\
$\sigma_{d}$ & $\hat{\pi}\hat{\sigma}_{12}$ & $-\varphi$\\
$\sigma_{d'} $ & $\hat{\pi}\hat{\sigma}_{23} $ & $-\varphi - \frac{2\pi}{3}$ \\
$\sigma_{d''} $ & $\hat{\pi}\hat{\sigma}_{31} $ & $-\varphi + \frac{2\pi}{3}$ \\
$C_6$ & $\hat{\pi}\hat{\sigma}_{231}$ & $\varphi + \frac{\pi}{3}$ \\
$C_6^{-1}$ & $\hat{\pi}\hat{\sigma}_{312}$ & $\varphi - \frac{\pi}{3}$ \\
\hline
\end{tabular}
\label{tab:elements}
\caption{The first column is the symmetry transformation designated by the cor\-re\-spond\-ing element of the point symmetry group of the regular hexagon permutation group ${\rm C}_{6v}$.  The second column is the same transformation expressed as the corresponding element of ${\rm S}_3 \times {\rm Z}_2$.  We use the notation for ${\rm S}_3$ permutation group elements such that $e$ is the identity, the 2-cycle $\hat{\sigma}_{ij}$ switches particles  $i$ and $j$, and 3-cycle $\hat{\sigma}_{ijk}$ switches $1$ to $i$, $2$ to $j$ and $3$ to $k$.  The element $\hat{\pi}$ is parity inversion.  The third column is the rotation or reflection matrix in Jacobi cartesian coordinates, and the fourth column is the equivalent transformation of the cylindrical Jacobi coordinate $\varphi$  Compare with Figure \ref{fig2}.}
\end{table}

For convenience, the irreducible representations of ${\rm C}_{6v}$ and the characters of the different classes of group elements are summarized in Table \ref{tab:c6v}.  These are useful for two reasons.  First, there will only be non-zero matrix elements of $V$ between vectors in the same type of irreducible representation.  This can be used to reduce the number of basis vectors that must be used to get results accurate up to a certain energy truncation $N_{max}$.  This simplification holds for the particular potential (\ref{eq:Vpolar}), but also holds for any two-particle interactions that depend only on the two-particle separation distances and have parity symmetry, for example Gaussian interactions or harmonic interactions.  Second, we can use these representation to implement the superselection rules.  For example, since the $\sigma_{v}$'s represent two-particle exchanges, the table shows that there are two one-dimensional representations that are bosonic under any exchange of particles, $A_1$ and $B_2$,  and two one-dimensional representations that are fermionic under any exchange of particles, $A_2$ and $B_1$.  The distinction between the $A$-type and $B$-type representations is the sign of the representation of parity inversion $\hat{\pi}\rightarrow C_2$.

\begin{table}
\centering
\begin{tabular}{|c|cccccc|}
\hline
$C_{6v}$ & $E$ & $C_2$ & $C_3$ & $C_6$ & $\sigma_v$ & $\sigma_d$\\
\hline
$A_1$ & 1 & 1 & 1 & 1 & 1 & 1\\
$A_2$ & 1 & 1 & 1 & 1 & -1 & -1\\
$B_1$ & 1 & -1 & 1 & -1 & -1 & 1\\
$B_2$ & 1 & -1 & 1 & -1 & 1 & -1\\
$E_1$ & 2 & -2 & -1 & 1 & 0 & 0\\
$E_2$ & 2 & 2 & -1 & -1 & 0 & 0\\
\hline
\end{tabular}
\caption{This the character table for the group ${\rm C}_{6v}$ which can be found in any reference on point symmetry groups and we include here for convenience.  The entries in the table are the character of the representation of the corresponding group element in each irreducible representation, i.e.\ the trace of the matrix that represents the transformation in any basis of the irreducible representation.  For example, the trace of the representation of the identity gives the dimension of the representation.}
\label{tab:c6v}
\end{table}

The Jacobi relative cylindrical basis vectors $\ckt{\nu,\mu}$ realized by the functions $\psi_{\nu,\mu}(\rho,\varphi)$ (\ref{eq:Jcylwave}) are not actually the basis vectors that transform irreducibly unless $\mu=0$ (which transforms under the trivial identity representation).  However, we can define the basis vectors
\begin{eqnarray}
\ckt{\nu,\mu,+} &=& \frac{1}{\sqrt{2}} \left(\ckt{\nu,\mu} + \ckt{\nu, -\mu}\right)\nonumber\\
\ckt{\nu,\mu,-} &=& \frac{1}{\sqrt{2}i} \left(\ckt{\nu,\mu} - \ckt{\nu, -\mu}\right)
\label{eq:pm}
\end{eqnarray}
for any $\mu>0$.  These are realized by the wave functions
\begin{eqnarray}
\psi^+_{\nu,\mu}(\rho,\varphi) = \sqrt{\frac{2\nu!}{\pi (\nu+\mu)!}}\rho^{\mu} L^{\mu}_{\nu}(\rho^2)\exp(-\rho^2/2)\cos(\mu\varphi)\nonumber\\
\psi^-_{\nu,\mu}(\rho,\varphi) = \sqrt{\frac{2\nu!}{\pi (\nu+\mu)!}}\rho^{\mu} L^{\mu}_{\nu}(\rho^2)\exp(-\rho^2/2)\sin(\mu\varphi)\label{eq:pmwave}
\end{eqnarray}
Using (\ref{eq:pmwave}) and the transformations given in Table \ref{tab:elements}, we see the vectors $\ckt{\nu,0}$, $\ckt{\nu,\mu,+}$ and $\ckt{\nu,\mu,-}$ form useful bases for the irreducible representations of ${\rm C}_{6v}$.  See Table \ref{tab:repbasis} for the exact correspondences.  For example, the basis vectors $\ckt{\nu,0}$ are elements of an $A_1$ bosonic, positive parity representation and these vectors are realized by cylindrically symmetric wave functions.  The $\ckt{\nu,6j,+}$ vectors (where $j$ is a non-negative integer) also transform under $A_1$, so they will mix with the $\ckt{\nu,0}$ states under the interaction $V$.  Vectors like $\ckt{\nu,6j,+}$ have wave functions with maximal magnitude at three-atom configurations (e.g.\ $\phi=0$, see Figure \ref{fig2}) and at atom-dimer configurations (e.g.\ $\phi=\pi/6$).  In contrast, any three-fermion vectors in $A_2$ and $B_1$ are realized by wave functions that have nodal lines in the relative plane at the atom-dimer configurations (and at the trimer configuration point at the origin).

\begin{table}
\centering
\begin{tabular}{|cc|c|c|c|}
\hline
$\mu$ & $\pm$ & $C_{6v}$ & $C_{2v}$ & Possibilities\\
\hline
0 & N.A. & $A_1$ & $A_1$ & BBB, BBX, XYZ \\
1& $+$ & $E_1$ & $B_1$ & FFX, XYZ \\
1& $-$ & $E_1$ & $B_2$ & BBX, XYZ \\
2& $+$ & $E_2$ & $A_1$ & BBX, XYZ \\
2& $-$ & $E_2$ & $A_2$ & FFX, XYZ \\
3& $+$ & $B_1$ & $B_1$ & FFF, FFX, XYZ \\
3& $-$ & $B_2$ & $B_2$ & BBB, BBX, XYZ \\
4& $+$ & $E_2$ & $A_1$ & BBX, XYZ \\
4& $-$ & $E_2$ & $A_2$ & FFX, XYZ \\
5& $+$ & $E_1$ & $B_1$ & FFX, XYZ \\
5& $-$ & $E_1$ & $B_2$ & BBX, XYZ \\
6& $+$ & $A_1$ & $A_1$ & BBB, BBX, XYZ \\
6& $-$ & $A_2$ & $A_2$ & FFF, FFX, XYZ \\
\hline
\end{tabular}
\caption{This table identifies the energy eigenbasis vectors $\ckt{\nu,0}$ for $\mu=0$ (defined in Section 2) and the energy eigenbasis vectors $\ckt{\nu,\mu,\pm}$ for $\mu>0$ defined in (\ref{eq:pm}) with their corresponding symmetry representations and superselection rules.  The pattern repeats for vectors with $\mu>6$.  BBB (FFF) means three identical bosons (fermions); BBX (FFX) two identical bosons (fermions) and one other particle; XYZ three distinguishable particles. }
\label{tab:repbasis}
\end{table}

The two two-dimensional representations $E_1$ and $E_2$ do not correspond to three indistinguishable particles.  As with all six representations, they could represent three distinguishable particles, but they also can represent two indistinguishable particles and a third distinguishable particle.  Assume labels 1 and 2 refer to the two indistinguishable particles and must be symmetrized or antisymmetrized.  The elements $\{E, \sigma_{v}, \sigma_{d}, C_2\}$ form a subgroup of ${\rm C}_{6v}$.  This point symmetry group ${\rm C}_{2v}\sim {\rm S}_2\times {\rm Z_2}$ consists of exchanges of only particles 1 and 2 and parity inversion and therefore can be used to classify superselection sectors for BBX and FFX configurations.  See Table \ref{tab:c2v} for the character table of this group.  The $E_1$ and $E_2$ irreducible representations of ${\rm C}_{6v}$ are reducible with respect to the subgroup ${\rm C}_{2v}$, leading to the further identifications in Table \ref{tab:repbasis}.

\begin{table}
\centering
\begin{tabular}{|c|cccc|}
\hline
$C_{2v}$ & $E$ & $C_2$ & $\sigma_v$ & $\sigma_d$ \\
\hline
$A_1$ & 1 & 1 & 1 & 1 \\
$A_2$ & 1 & 1 & -1 & -1\\
$B_1$ & 1 & -1 & -1 & 1\\
$B_2$ & 1 & -1 & 1 & 1\\
\hline
\end{tabular}
\caption{This the character table for the four, one-dimensional ir\-re\-duc\-ible re\-presentations of the group ${\rm C}_{2v}$.}
\label{tab:c2v}
\end{table}

We should note again that we are assuming that there is no entanglement between the spatial wave function and any internal degrees of freedom like spin or hyperfine energy level.  Additionally, we are assuming that there is not entanglement among the internal degrees of freedom themselves.  These assumptions imply that these three equal-mass particles all have well-defined, separable internal states.  When there are three different internal states, we call the configuration XYZ.  If two internal states are populated, or it is a mixture of bosons and fermions, it is either BBX or FFX.  Relaxing these assumptions about separability, the possible state space for various particle configurations become broader.  For example, one could place three fermions into superposition of three distinguishable internal states that is antisymmetric under pairwise exchange.  Then those three fermions could have a spatial wave function that was totally symmetric.  The possibilities are further extended when entanglement between internal state and spatial wave function is considered.

\section{Results in truncated Hilbert space}

In order to find the approximate eigenstates of the full Hamiltonian $H=H_0+V$, we calculate the matrix elements of the interaction $V$ in the basis states of $H_0$.  The cylindrical Jacobi basis states defined in the last section exploit the symmetries of $H_0$, $V$ and $H$ so that, for a given energy level truncation, the fewest number of matrix elements need to be calculated.  Of course a large improvement occurs simply because the full Hamiltonian retains its center-of-mass separability.  For a given maximum $N$, center-of-mass/relative separability reduces the order of the model system from $1/6(N+3)(N+2)(N+1)\sim O(N^3)$ to $1/2(N+2)(N+1)\sim O(N^2)$ because the effective one-body center-of-mass problem is solvable.  Additionally, the decomposition of the relative space into irreducible blocks means that matrix elements between vectors in different representations are zero, further reducing the size of the computation.  Although this representation reduction does not change the order of the computation from $O(N^2)$, it does reduce the size of computation needed for a given truncation size $N$.  For example, for three identical particles with either parity, only one out of every twelve states contribute to the calculation.  To give a flavor of this, Table \ref{tab:deg3} shows the reduction of the dimension of the truncated Hilbert space for three identical bosons and fermions for $N=0$ to $N=12$.

\begin{table}
\centering
\begin{tabular}{|c|c|c|c|c|c|c|}
\hline
$N$ & $D_{N}$& $D_{\tilde{N}}$ & $D^{\rm 3B+}_{\tilde{N}}$ & $D^{\rm 3B-}_{\tilde{N}}$ & $D^{\rm 3F+}_{\tilde{N}}$ & $D^{\rm 3F-}_{\tilde{N}}$  \\
\hline
0 & 1 & 1 & 1 & 0 & 0 & 0 \\
1 & 4 & 3 & 1 & 0 & 0 & 0 \\
2 & 10 & 6 & 2 & 0 & 0 & 0 \\
3 & 20 & 10 & 2 & 1 & 1 & 0 \\
4 & 35 & 15 & 3 & 1 & 1 & 0 \\
5 & 56 & 21 & 3 & 2 & 2 & 0 \\
6 & 84 & 28 & 5 & 2 & 2 & 1 \\
7 & 120 & 36 & 5 & 3 & 3 & 1 \\
8 & 165 & 45 & 7 & 3 & 3 & 2 \\
9 & 220 & 55 & 7 & 5 & 5 & 2 \\
10 & 286 & 66 & 9 & 5 & 5 & 3 \\
11 & 364 & 78 & 9 & 7 & 7 & 3 \\
12 & 455 & 91 & 12 & 7 & 7 & 5 \\
\hline
\end{tabular}
\caption{The total excitation number $N$ identifies the energy eigenspace with $E_N = \hbar \omega (N +3/2)$.  The dimension of the truncated Hilbert space up to and including that energy level is $D_N = 1/6(N+1)(N+2)(N+3)$.  We define $D_{\tilde{N}}$ as the dimension of the truncated Hilbert space with maximum excitation $N$ but constrained such that center-of-mass motion is in the ground state.  In other words, for the Jacobi cylindrical basis, $D_{\tilde{N}}$ is the dimension of all relative states of distinguishable particles such that $\eta=0$ and $2\nu + |\mu| \leq N$.  The other dimensions are: $D^{\rm 3B+}_{\tilde{N}}$ for three identical bosons with positive parity, $D^{\rm 3B-}_{\tilde{N}}$ for three identical bosons with negative parity, and similar definition with fermions for $D^{\rm 3F+}_{\tilde{N}}$ and $D^{\rm 3F-}_{\tilde{N}}$.}
\label{tab:deg3}
\end{table}

We calculate the matrix elements in the cylindrical Jacobi basis using their realizations in relative coordinate space for $\mu=0$ with the wave function  $\psi_{\nu,0}(\rho,\varphi)$ defined in (\ref{eq:Jcylwave}) and for $\mu>0$ with $\psi^+_{\nu,\mu}(\rho,\varphi)$ and $\psi^-_{\nu,\mu}(\rho,\varphi)$ defined in (\ref{eq:pmwave}).  The coordinate representations of the wave functions are separable in $\rho$ and $\varphi$, and we exploit that to make the following definitions:
\begin{eqnarray}
\cbr{\nu', 0}V\ckt{\nu, 0} &=& \frac{g}{\sqrt{2}\sigma} R(\nu',0,\nu,0)A^{++}(0,0)\nonumber\\
\cbr{\nu', 0}V\ckt{\nu, \mu,+} &=& \frac{g}{\sigma} R(\nu',0,\nu,\mu)A^{++}(0,\mu)\nonumber\\
\cbr{\nu', \mu',\pm}V\ckt{\nu, \mu,\pm} &=& \frac{\sqrt{2}g}{\sigma}R(\nu',\mu',\nu,\mu)A^{\pm\pm}(\mu',\mu).
\label{eq:sepV}
\end{eqnarray}
The first two equations in (\ref{eq:sepV}) are only relevant for the $A_1$ representation of ${\rm C}_{6v}$ associated with superselection rules for the cases BBB, BBX or XYZ.  The changing factors of $\sqrt{2}$ are due to the normalization for the $\mu>0$ states defined in (\ref{eq:pmwave}).

The functions $A^{\pm\pm}(\mu',\mu)$ are symmetric in the $(\mu',\mu)$ argument and can be calculated directly from the integral
\begin{equation}
A^{\pm\pm}(\mu',\mu) = \sum_{i=1}^6 \int_0^{2\pi}d\varphi\ \delta(\varphi -  {\scriptstyle\frac{(2i-1)}{6}}\pi) \cdot \left\{\begin{array}{c} \cos(\mu'\varphi) \\ \sin(\mu'\varphi)\end{array} \right\}\cdot\left\{\begin{array}{c} \cos(\mu\varphi) \\ \sin(\mu\varphi) \end{array}\right\}\label{eq:Aint}
\end{equation}
where each $+$ corresponds to a cosine terms and each $-$ corresponds to a sine term. See Table \ref{tab:A} for a summary of the results of this integral. Explicit calculation shows that $A^{\pm\pm}(\mu',\mu)$ will only be non-zero when the particular combinations of $\{\mu,\pm\}$ and $\{\mu,\pm\}'$ come from the same irreducible representations of ${\rm C}_{6v}$.  Additionally, we see that states in representations $A_2$ and $B_1$ do not feel the two-body interaction potential at all, as expected for identical fermions.  Wave functions in these sectors all have nodal lines corresponding to configurations where two particles overlap, and are therefore antisymmetric under reflection across this line~\footnote{Note: for two-particle interactions with a finite range but still parity-symmetric (for eaxmple, Gaussian interactions or powers of $|q_i - q_j|$) the integrals to calculate $A^{\pm\pm}$ and $R$ will generally be more complicated, but the $A^{\pm\pm}$ for the fermion sectors will still be zero.  This is a consequence of this antisymmetry about atom-dimer configuration rays.}.  Therefore, the three indistinguishable fermionic relative states of the form $\ckt{\nu,6j+3,+}$ and $\ckt{\nu,6j,-}$ are eigenstates of both $H_0$ and $H$.

\begin{table}
\centering
\begin{tabular}{|c|c|l|l|}
\hline
$C_{6v}$ & $C_{2v}$ & XYZ$\pm$ & Matrix elements\\
\hline
$A_1$ & $A_1$ & BBB+ & $A^{++}(6j,6k) = 6(-1)^{j+k}$ \\
$A_2$ & $A_2$ & FFF+ & $A^{--}(6j,6k) = 0$\\
$B_1$ & $B_1$ & FFF- & $A^{++}(6j+3,6k +3 ) = 0$\\
$B_2$ & $B_2$ & BBB- & $A^{--}(6j+3,6k +3 ) = 6(-1)^{j+k}$\\
$E_1$ & $B_1$ & FFX- & $A^{++}(6j\pm 1,6k \pm 1) = 3(-1)^{j+k}$ \\
$E_1$ & $B_2$ & BBX- & $A^{--}(6j+3 \pm 2,6k +3 \pm 2) = 3(-1)^{j+k}$ \\
$E_2$ & $A_1$ & BBX+ & $A^{++}(6j\pm 2,6k \pm 2) = 3(-1)^{j+k}$ \\
$E_2$ & $A_2$ & FFX+ & $A^{--}(6j+3 \pm 1,6k +3 \pm 1) = 3(-1)^{j+k}$ \\
\hline
\end{tabular}
\caption{This table summarized the results of the angular integral (\ref{eq:Aint}) for the different irreducible representations of $C_{6v}$ and $C_{2v}$.  The letters of the third column describe the `most indistinguishable' three particle configuration with that representation and the corresponding parity.  For example, ``BBB+'' in the first line can also correspond to BBX and XYZ states with positive parity.}
\label{tab:A}
\end{table}

To analyze states in the other four irreducible representations, we must calculate the radial integral:
\begin{equation}
R(\nu',\mu',\nu,\mu) =  \frac{\sqrt{\nu!\nu'!}}{\pi\sqrt{(\nu+\mu)!(\nu'+\mu')!}} \int_0^\infty \!\!\!\!d\rho\,\rho^{\mu + \mu'+1}L^\mu_\nu\! \left(\rho^2\right)\! L^{\mu'}_{\nu'}\!\left(\rho^2\right) e^{-\rho^2}.
\label{eq:r1}
\end{equation}
Using the substitution $\rho^2 = t$, this can be put into the standard form
\begin{equation}
R(\nu',\mu',\nu,\mu) = \frac{\sqrt{\nu!\nu'!}}{2\pi\sqrt{(\nu+\mu)!(\nu'+\mu')!}} \int_0^\infty  \!\!\!\!dt\, t^{\frac{\mu + \mu'-1}{2}} L^\mu_\nu\left(t\right) \! L^{\mu'}_{\nu'}\left(t\right) e^{-t}
\label{eq:r2}
\end{equation}
which can be evaluated as~\cite{mathematica}
\begin{eqnarray}
R(\nu',\mu',\nu,\mu) &=&  \frac{1}{2\pi}\sqrt{\frac{(\nu'+\mu')!}{\nu!\nu'!(\nu+\mu)!}} \frac{\Gamma\left(\nu + {\scriptstyle \frac{1 + \mu - \mu'}{2}}\right)\Gamma\left( {\scriptstyle \frac{1 + \mu + \mu'}{2}}\right)}{\Gamma\left( {\scriptstyle \frac{1 + \mu - \mu'}{2}}\right)}\\
&&\times  {}_3F_2\left(-\nu' ,{\scriptstyle \frac{1+\mu+\mu'}{2}},{\scriptstyle \frac{1-\mu+\mu'}{2}};-\nu+{\scriptstyle \frac{1-\mu+\mu'}{2}},\mu'+1;1\right).\nonumber
\label{eq:r3}
\end{eqnarray}
The function ${}_3F_2(\cdots;\cdot\,\cdot;\cdot)$ is a generalized hypergeometric function.  The lack of symmetry between primed and unprimed coordinates in (\ref{eq:r3}) is deceiving.  One can either use the definition (\ref{eq:r1}) or the properties of ${}_3F_2(\cdots;\cdot\,\cdot;\cdot)$ to show that $R(\nu',\mu',\nu,\mu)=R(\nu,\mu,\nu',\mu')$.

\begin{figure}
\centering
\includegraphics[width=\linewidth]{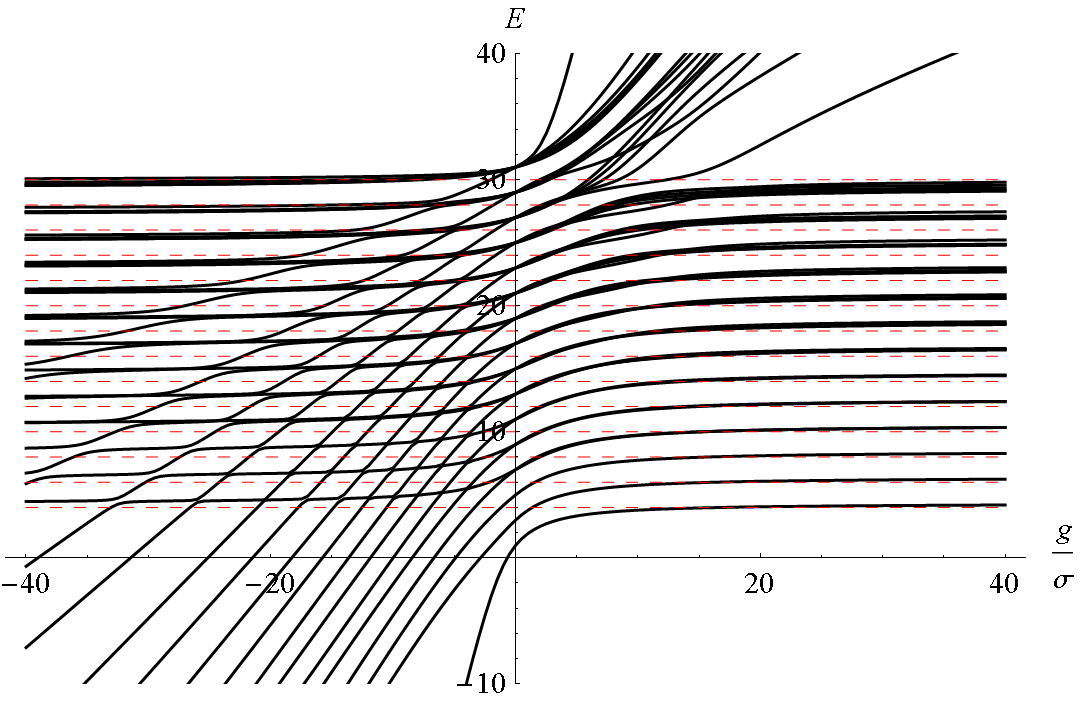}
\caption{(Color online) Solid black lines are the 40 energy levels with symmetry corresponding to the totally symmetric, positive parity $A_1$ representation of ${\rm C}_{6v}$ for $\tilde{N}_{max}=30$.  Energies are measured in units of $\hbar \omega$. The dashed red lines are the exact result for universal states in the limit $g/\sigma \rightarrow\infty$ and correspond to the energy levels and degeneracies of the non-interacting FFF+ $A_2$ representation.  This representation contains the ground state of the $BBB$, $BBX$ and $XYZ$ configurations that is a trimer in the case of attractive interactions.  See text for more details.   }
\label{fig:bosonA} 
\end{figure}

\begin{figure}
\centering
\includegraphics[width=\linewidth]{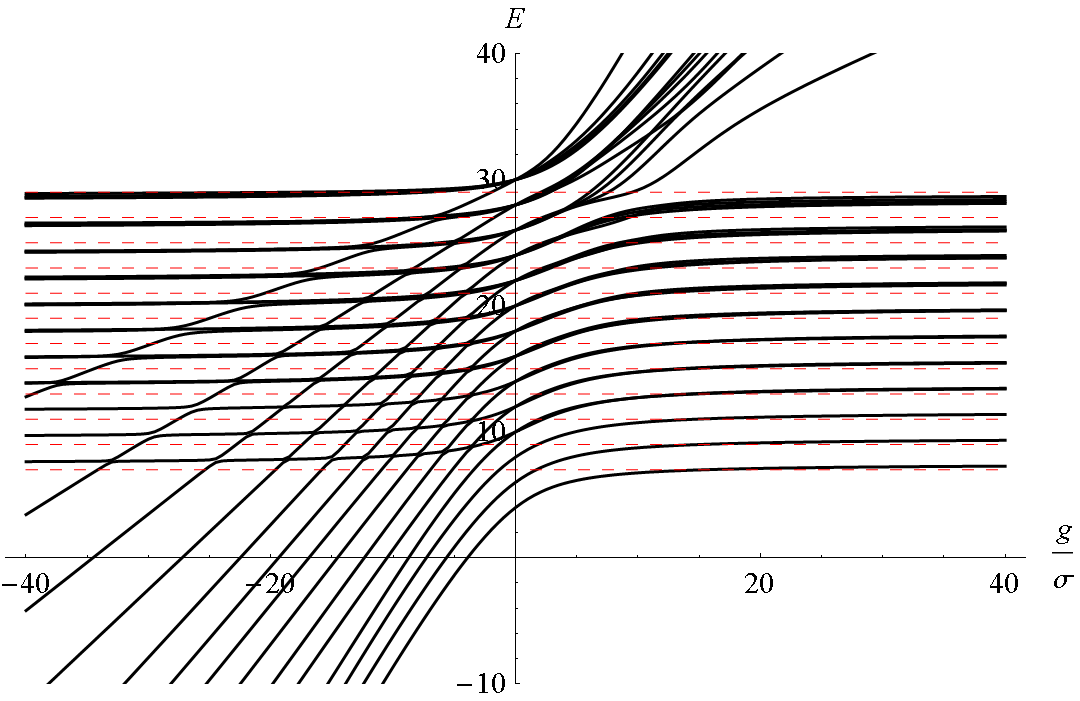}
\caption{(Color online) Solid black lines are the 40 energy levels with symmetry corresponding to the totally symmetric, negative parity $B_2$ representation of ${\rm C}_{6v}$ for $\tilde{N}_{max}=30$.  Energies are measured in units of $\hbar \omega$. The dashed red lines are the exact result for universal states in the limit $g/\sigma \rightarrow$ and correspond to the energy levels and degeneracies of the non-interacting FFF- $B_1$ representation.}
\label{fig:bosonB} 
\end{figure}

Combining the results of Table \ref{tab:A} and equation (\ref{eq:r3}) to calculate the matrix elements (\ref{eq:sepV}), we can proceed with some numerical results.  Figures \ref{fig:bosonA} and \ref{fig:bosonB} depict the relative energy levels calculated up to $\tilde{N}_{max}=2\nu + \mu=30$ (i.e. $E^{rel}_{max} = 31 \hbar\omega$) for the two indistinguishable boson sectors.  This corresponds to a 51-state basis for the positive-parity bosonic sector $A_1$ and  40 state basis for the bosonic $B_2$ sector. Note that in these energy level diagrams, all apparent crossings are actually avoided crossings.  Incorporating center-of-mass motion only adds a shift of $\hbar \omega(\eta+1/2)$ to each of these levels.

Looking at Figure \ref{fig:bosonA}, three different kinds of states are identifiable.  First, the states whose energies are are approximately horizontal for $g/\sigma \approx E^{rel}_{max}$ are three-atom states.  This can be most easily seen for the low energy states, where as $|g/\sigma|$ becomes much larger than $\hbar \omega$ the bosonic states become effectively fermionic, i.e.\ they approach the exact energy and degeneracy expected for the non-interacting, postive-parity three-fermion $B_1$ sector.   The wave functions for these states all have maximum amplitudes in the sectors of relative space corresponding to three disparate particles, specifically $\varphi=0$, $\pm\pi/3$, and $\pm 2\pi/3$.  In other words, in these spectra we can observe the Tonks-Girardeau phenomenon of fermionization of bosons with strong interactions~\cite{girardeau_relationship_1960,lieb_exact_1963-1,lieb_exact_1963}.   These states can be considered universal in the unitary limit of $|g/\sigma|\gg \hbar \omega$ because the energy levels only depend on the trap parameters and not the interaction~\cite{braaten_universality_2006,werner_unitary_2006}.  For higher energy states near the cutoff, the numerical agreement between the bosonic state and the corresponding fermionic limit becomes less precise, although the bands still have the correct degeneracy.  This systematic effect is due to the truncation of the Hilbert space, and occurs even though the interaction potential is renormalizable in one-dimension.  Methods from nuclear physics exist for transforming the Hamiltonian into an effective Hamiltonian to account for this effect, e.g. the Block-Horowitz approach used in \cite{luu_three-fermion_2007} or the Lee-Suzuki method from \cite{navratil_few-nucleon_2000}, and these will be explored in subsequent work.

For $g<0$, the three-atom `gaseous' states would not be stable in Figure \ref{fig:bosonA}, but could decay into `liquidlike/solidlike' states~\cite{blume_three_2002} consisting of the two other types of states: trimer states and atom-dimer states.  The trimer state only occurs in the $A_1$ sector because only $\mu=0$ states can have non-zero probabilities amplitudes for all three particles to be in the same location. Above this state are the atom-dimer states, which diverge as $g$ decreases from zero and the dimer becomes more tightly bound.  For $|g/\sigma| \lesssim E^{rel}_{max}$, the energy eigenvalue diverges roughly quadratically, and the difference between the different atom-dimer states is just the energy cost of exciting the atom and/or dimer to a greater value of $\nu$.  For $|g/\sigma| \gtrsim  E^{rel}_{max}$, the atom-dimer and trimer states appear to diverge linearly, but this is again a limitation of the truncated Hilbert space.  As one increases the number of basis vectors in this section, the expected quadratic divergence holds for larger values of of the interaction parameter $g$.  Additionally, more atom-dimer states appear and the spacing between those levels for a given $g<0$ becomes more exactly $2\hbar\omega$.  

A couple other interesting features are the stepped cascades of gaseous states for increasing negative values of $g$ and the divergent states with $g>0$.  The divergent states for $g>0$ are again an artifact of the truncation.  As the cutoff is raised and additional energy levels are added, the lowest divergent states bend down to the fermionic limit.  On the other hand, the cascade structure is not an artifact of the truncation.  As higher energy levels are added, the steps of the cascading energy level become more regular for low negative $g$.  These structures depends on coupling between $\mu$ states of different relative (configuration space) angular momenta and are absent when that coupling is removed.

\begin{figure}
\centering
\includegraphics[width=\linewidth]{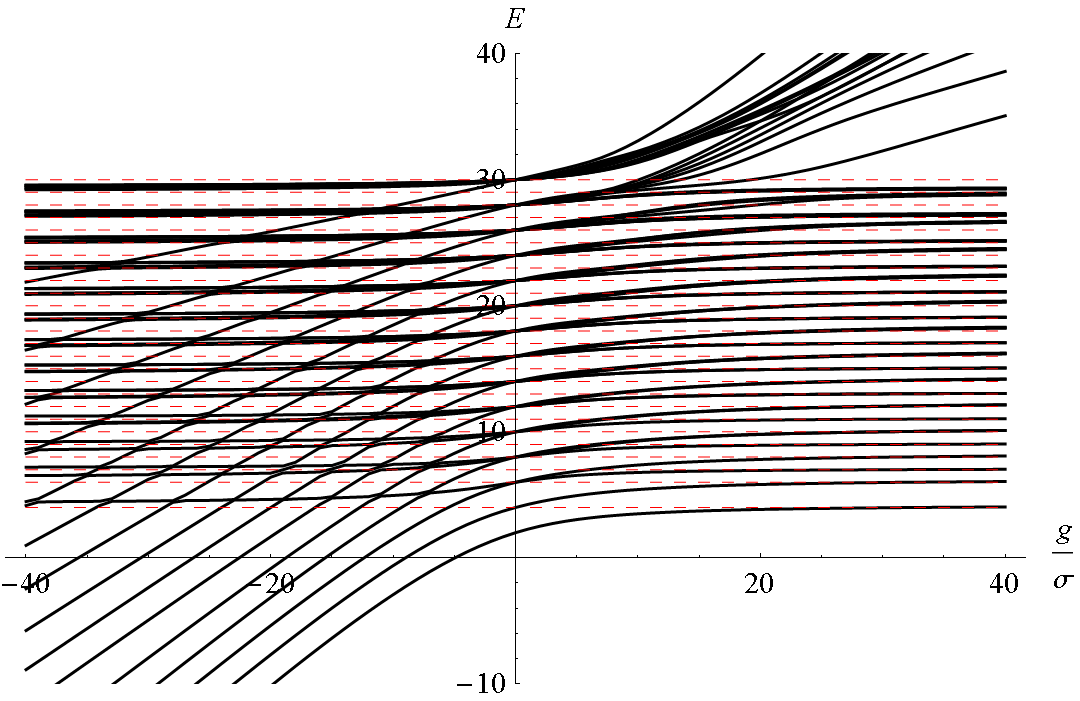}
\caption{(Color online) Solid black lines are the 80 energy levels with symmetry corresponding to the two-component fermion with negative parity in the $E_1$/$B_2$ representation of ${\rm C}_{6v}$/${\rm C}_{2v}$ for $\tilde{N}_{max}=30$.  Energies are measured in units of $\hbar \omega$. This is the ground state sector for the two-component fermion.  The dashed red lines are the exact result for universal states for both the FFF+ $A_2$ and the FFF- $B_1$ representations, showing the emergence of energy levels corresponding to `anomalous' parity states in the unitary limit.}
\label{fig:FFXE1B1} 
\end{figure}

As a final comparison, we also include Figure \ref{fig:FFXE1B1} which depicts the spectrum for the $E_1$/$B_1$ representation, corresponding to an FFX configuration, such as two spin-up fermions and one spin-down fermion.

\section{Concluding Remarks}

Some extensions to more particles of the method described here are straightforward.  For example, the representation theory of permutation groups in three dimensions is well-known, and an extension to four particles in one-dimension can be accomplished using the octahedral symmetry of the relative configuration space.  Further extensions to more than five particles in one dimension require using the less-familiar (at least from molecular and solid state physics) point groups in higher dimensions, which have been classified by Coxeter and others.  Since the three-body problem in one-dimension carries so many properties of the three-dimensional system, such as universality and bound state structure, this line of investigation seems promising.

When extending this method to higher particle numbers \emph{and} higher dimensions, for example $N$ particles in $d$ dimensions, one difficulty is that the realization of the permutation group ${\rm S}_N$ in ${\rm O}((N-1)d)$ no longer commutes with the ${\rm O}((N-1)d)$-representation of the ${\rm O}(d)$ symmetry of the two-particle interaction, a symmetry that is also no longer a discrete symmetry.  Contrast this with the present case:  the ${\rm O}(2)$ realization of ${\rm S}_3$ in the Jacobi plane commutes with the ${\rm O}(2)$ parity representation of the discrete ${\rm Z}_2 \equiv {\rm O}(1)$ symmetry of the interaction.  However, we do not find this task hopeless, and note that even in $d$-dimensions, there is always ${\rm O}((N-1)d)$ representations of particle permutations.  Such geometric realizations of the symmetries of the relative configurations of three particles appear in the construction of Talmi-Moshinksy brackets used in \cite{tolle_spectrum_2010,tolle_universal_2011} and the perturbative expansion solution of Faddeev equations in \cite{stoll_production_2005}.  

Additionally, in higher dimensions there are the difficulties of renormalization of the zero-range interaction, but methods from effective field theory and/or regularization have previously been used to study these interactions in truncated Hilbert spaces and their results should be extensible.  Similar approaches will also be necessary if genuine multi-body interactions are included; for example, a term proportional to $\delta(\rho)$ in the three-body, one-dimensional case.

For anisotropic traps, the rotational symmetry is gone, but there can still be reflection symmetries whose representations can be exploited for constructing symmetry adapted bases using these methods, as shown by \cite{grishkevich_theoretical_2011}.  Similarly, even with particle of uneven masses, discrete symmetries can help reduce the problem in a numerically more efficient way.

As a final comment, the same symmetry techniques that provide efficiency through separability and reducibility in the present work should also be useful for calculating the entanglement among atoms using truncated Hilbert spaces.  The coefficients that connect the Jacobi cylindrical basis can be calculated using a variety of symmetry-based methods.  In the case of indistinguishable bosons, these coefficients can be used to find the distance from the energy eigenstate to a separable state as quantified by any atom-dimer bipartition.  In the case of indistinguishable fermions, the entanglement can be compared to the least entanglement one can expect in an antisymmetrized state.  In either case, we believe that entanglement spectroscopy of trapped few-body systems could be an interesting testing ground for the interplay of particle superselection roles and the emergence of composite systems.

\section*{Acknowledgments}

The author would like to thank J. Uscinski, M. Roberts, B. Weinstein, A. Taylor, J. Revels, and especially P.R. Johnson for numerous discussions.

\end{document}